\documentstyle[12pt]{article}
\title{The Spacetime View of the Information Paradox}
\author{Hrvoje Nikoli\'c \\
Theoretical Physics Division, Rudjer Bo\v{s}kovi\'{c} Institute, \\
P.O.B. 180, HR-10002 Zagreb, Croatia \\
{\normalsize e-mail: hrvoje@thphys.irb.hr} \\
\makebox[1in]{} \\
}
\date{\today}
\begin{document}
\maketitle
\begin{abstract}
In semiclassical gravity, the final state of black-hole evaporation
cannot be described by a pure state. Nevertheless, 
we point out that the system can be described
by a {\em generalized} pure state, which is not defined
on a 3-dimensional hypersurface but on the 4-dimensional spacetime
describing the whole Universe at all times. Unlike the conventional
quantum state, such a generalized state treats time on an equal footing with space,
which makes it well suited for systems that are both
quantum and relativistic. In particular,
such a generalized state contains a novel type of information encoded in the correlations
between future and past, which avoids the black-hole information paradox.
\end{abstract}
\vspace*{0.5cm}
{\it Keywords}: quantum information; spacetime; black-hole information paradox 
\vspace*{0.9cm} 

\section{Introduction}

Semiclassical gravity predicts that black holes evaporate \cite{hawk1}
and that the final state after the complete evaporation cannot be described
by a pure state \cite{hawk2}, making the time evolution of the system 
non-unitary. Various attempts to restore a unitary time evolution have been
proposed (see, e.g., \cite{gid,har,pres,pag,gid2,str,math,hoss} for reviews), 
but none of them seems completely satisfying.
On the other hand, the idea that {\em fundamental} dynamics should allow
time evolution from a pure to a non-pure (i.e., mixed) state \cite{hawk2} is usually discarded
due to serious pathologies that occur \cite{banks}.
(By emphasizing the word ``fundamental'' we stress that this does not refer
to non-fundamental emergent evolution from pure to mixed states caused by
environment-induced decoherence, which, of course, does not lead to
pathologies of the kind discussed in \cite{banks}.)

In this paper we present a simplified explanation of a novel way out of this problem
recently proposed in \cite{nik}. (A qualitatively similar idea based on a
completely different formalism has also been introduced in \cite{hartle1,hartle2}.)
In this paper we put emphasis on the quantum-information aspects of the proposal,
in a manner that requires only rudimentary familiarity with concepts of general
relativity and black hole physics.

More specifically, we show that it is possible to allow the possibility that state at a given time cannot be described
by a pure state, and yet to avoid pathological dynamics describing evolution from pure to mixed
states. We do that by avoiding the concept of ``time evolution'' itself, or more precisely
by replacing the concept of time evolution with something more general. While a generalization of
the concept of time evolution may seem unjustified at first sight, we point out that such a generalization
emerges naturally and almost trivially from the requirement that time in quantum theory 
should be treated on an equal footing with space. The latter requirement hardly needs additional
justification, because treating time on an equal footing with space lies in the root of
the theory of relativity, both special and general. 

In Sec.~\ref{SEC2} we explain in more detail how time in 
quantum mechanics can be treated on an equal footing with space,
while the application to black-hole information paradox is discussed
in Sec.~\ref{SEC3}. In Sec.~\ref{SEC4} we qualitatively discuss some generalizations
and draw the conclusions.

\section{Quantum information in spacetime}
\label{SEC2}

The basic idea is simple. An $n$-particle state is usually described by a wave function
of the form $\psi({\bf x}_1,\ldots ,{\bf x}_n,t)$. Clearly, in this usual description
time is not treated on an equal footing with space; there are $n$ space coordinates ${\bf x}_a$,
$a=1,\ldots n$, while there is
only one time coordinate $t$. A natural thing to do is to introduce a many-time formalism 
\cite{dirac,rosenfeld,dfp,bloch,tomon}, 
in which the wave function generalizes to $\psi({\bf x}_1,t_1\ldots ,{\bf x}_n,t_n)$.
Then the usual single-time wave function is only a special case defined by
\begin{equation}\label{E1}
 \psi({\bf x}_1,\ldots ,{\bf x}_n,t) =
\psi({\bf x}_1,t_1\ldots ,{\bf x}_n,t_n) |_{t_1=\cdots =t_n=t} .
\end{equation}
Introducing the relativistic notation $x=({\bf x}, t)$, the many-time wave function can be written in a
manifestly relativistic form $\psi(x_1, \ldots ,x_n)$.
While the single-time wave function $\psi({\bf x}_1,\ldots ,{\bf x}_n,t)$ describes
the correlations between particles at a common time, the many-time wave function
$\psi(x_1, \ldots ,x_n)$ describes the correlations between particles at different times.
The separation between different points $x_a$ does not even need to be spacelike,
but may be null or timelike as well.
Therefore, unlike the usual single-time wave function, the many-time wave function may describe
the correlations between future and past. 

The probabilistic interpretation of wave functions can also be generalized 
such that time is treated on an equal footing with space \cite{nikijqi}.
The (infinitesimal) probability that $n$ particles will be found
in an infinitesimal configuration-space volume $d^4x_1\cdots d^4x_n$ 
around the points $x_1, \ldots ,x_n$ is equal to
\begin{equation}\label{E2}
 dP=|\psi(x_1, \ldots ,x_n)|^2 d^4x_1\cdots d^4x_n .
\end{equation}
(More precisely this is valid in Minkowski spacetime, while in curved spacetime
one replaces $d^4x$ with $d^4x \, |g(x)|^{1/2}$, where $g(x)$ is the determinant
of the metric tensor.)
This is the {\it a priori} probability, describing 
the situation in which nothing, except the wave function $\psi(x_1, \ldots ,x_n)$, 
is known about the spacetime positions of detected particles.
On the other hand, if one {\em knows} that all particles are detected at the 
specific time $t_1=\cdots =t_n=t$, then the {\it a priori} probability should be replaced
by the {\em conditional} probability. The {\it a priori} probability (\ref{E2})
implies the conditional probability
\begin{equation}\label{E3}
dP_{(3n)}=\frac{|\psi({\bf x}_1,\ldots,{\bf x}_n,t)|^2 d^3x_1 \cdots d^3x_n}
{N_{t}} ,
\end{equation}
where 
\begin{equation}\label{e8n}
N_{t}=\int |\psi({\bf x}_1,\ldots,{\bf x}_n,t)|^2 d^3x_1 \cdots d^3x_n 
\end{equation}
is the normalization factor.
Thus, the usual familiar probabilistic interpretation in space (\ref{E3})
is nothing but a special case of the general
probabilistic interpretation in spacetime (\ref{E2}).

Note that (\ref{E2}) implies that the time-coordinate is also treated as a quantum variable,
in the same sense as the space-coordinate. This has the following geometrical interpretation.
While the usual wave function $\psi({\bf x}_1,\ldots,{\bf x}_n,t)$
with $t$ treated as an external parameter represents a state on a given 3-dimensional
spacelike hypersurface having a constant $t$, {\em the generalized wave function
$\psi(x_1, \ldots ,x_n)$ interpreted according to (\ref{E2}) 
represents a state on the whole 4-dimensional spacetime describing the Universe at all times.}

The above can also be generalized to quantum field theory (QFT), in which the number of particles
may be uncertain. An arbitrary QFT state can be written as a superposition of states with different
definite numbers $n=0,1,2,3, \ldots$ of particles. In this way, a QFT state can be represented by a wave function $\psi(x_1,x_2,\ldots)$ 
depending on an infinite number of coordinates $x_a$ \cite{nikqft}. 

\section{Application to black holes}
\label{SEC3}

Now it is easy to understand how the black-hole information paradox gets resolved.
For simplicity, we consider a single pair of  entangled particles, where one particle is inside and the
other particle is outside 
of the black hole. (A more realistic treatment with an uncertain number of particles
is also possible \cite{nik}.)
The wave function of the inside particle can be expanded in terms of modes 
$\{ \psi^{\rm(in)}_l(x_{\rm in}) \}$ (see the Appendix).
We assume that the black hole, here treated as a classical background, evaporates completely
at time $t=t_{\rm evap}$. Therefore the inside particle does not exist for $t>t_{\rm evap}$, or 
more precisely the probability of finding the inside particle in the region $t>t_{\rm evap}$
vanishes. Consequently,
\begin{equation}\label{E5}
\psi^{\rm(in)}_l({\bf x}_{\rm in},t_{\rm in}) |_{t_{\rm in}>t_{\rm evap}}= 0 .
\end{equation}
The wave function of the outside particle can be expanded in terms of modes
$\{ \psi^{\rm(out)}_k(x_{\rm out}) \}$ (see the Appendix),
so the most general wave function describing an entanglement between 
the inside particle and the outside particle has a form
\begin{equation}\label{E6}
\psi(x_{\rm in}, x_{\rm out})=\sum_{l}\sum_{k}c_{lk}\psi^{\rm(in)}_l(x_{\rm in}) \psi^{\rm(out)}_k(x_{\rm out}) .
\end{equation}
This pure state can also be described by the pure density matrix
\begin{equation}\label{E7}
 \rho(x_{\rm in}, x_{\rm out}|x'_{\rm in}, x'_{\rm out})=
\psi(x_{\rm in}, x_{\rm out}) \psi^*(x'_{\rm in}, x'_{\rm out}) .
\end{equation}
The outside state is described by the mixed state
\begin{equation}\label{E8}
 \rho^{\rm(out)}(x_{\rm out}|x'_{\rm out})=\int d^4x_{\rm in}\, |g(x_{\rm in})|^{1/2} \,
\rho(x_{\rm in}, x_{\rm out}|x_{\rm in}, x'_{\rm out}) ,
\end{equation}
where the factor $|g(x_{\rm in})|^{1/2}$ is incorporated because
we work in curved spacetime. 
From (\ref{E6}) and (\ref{E7})
we see that the system can be described by a pure state even after the complete evaporation.
However this description is trivial, because (\ref{E5}) implies that (\ref{E6}) and (\ref{E7})
vanish for $t_{\rm in}>t_{\rm evap}$. 
Physically, this vanishing expresses the fact that the inside particle cannot be observed
(i.e., the probability amplitude is zero) at times after the evaporation, no matter when or where
the outside particle is observed. This is nothing but a way to express the fact that
the inside particle is destroyed, i.e., does not longer exist after the complete evaporation.
Still, even a nontrivial pure-state description for 
$t_{\rm out}>t_{\rm evap}$ is possible, provided that $t_{\rm in}$ is restricted to the
region $t_{\rm in}<t_{\rm evap}$. In this case (\ref{E6}) and (\ref{E7}) describe the
correlations between the outside particle after the complete evaporation and 
the inside particle before the complete evaporation. In other words, if one asks where the 
information after the complete evaporation is hidden, then the answer is -- 
{\em it is hidden in the past}.
Of course, experimentalists cannot travel to the past, so information is lost for the experimentalists.
Yet, this information loss is described by a pure state, so one does not need to use the
Hawking formalism \cite{hawk2} in which a state evolves from a pure to a mixed state.
By avoiding this formalism one avoids its pathologies \cite{banks} too, which may be 
viewed as the main advantage of our approach.

\section{Discussion and conclusion}
\label{SEC4}

The analysis in Sec.~\ref{SEC3} was based on semiclassical gravity,
but the main idea can also be incorporated beyond semiclassical gravity.
The main requirement is to have
many time-degrees of freedom. This requirement is achieved, e.g., by
canonical approaches to quantum gravity, including loop quantum gravity \cite{rov}. 
In such approaches
the state satisfies the Hamiltonian constraint ${\cal H}\psi[g,\phi]=0$, where
$g$ and $\phi$ represent the gravitational and matter degrees of freedom, respectively. Some of the matter degrees of freedom are interpreted as the ``clock'' time, so we write $\phi=(\phi',T)$. 
The clock-time variable $T$ may contain many local degrees of freedom, with a different time
for degrees of freedom inside and outside of the black hole.
This leads to a resolution of the black-hole information paradox similar 
to that in Sec.~\ref{SEC3} \cite{niktimeqg}.

String theory can also incorporate the ideas of Secs.~\ref{SEC2} and \ref{SEC3}.
Instead of a QFT state represented by a wave function
$\psi(x_1,x_2,...)$ with an infinite but discrete set of coordinates $x_a$, a state in string theory
is represented by a wave functional $\psi[x(\sigma)]$ where the discrete label
$a$ is replaced by a continuous label $\sigma$. By allowing $x^{\mu}(\sigma)$ to be a discontinuous function, the wave functional $\psi[x(\sigma)]$ may describe not only 1-string states, but many-string
states as well \cite{nikstr}.

In short, whenever the quantum theory is formulated such that time is a {\em local} variable
quantized as one of the degrees of freedom, then time inside the black hole is not the same
variable as time outside of the black hole. Such systems can be completely described by pure states
even when not all information is available at a given spacelike hypersurface. 
Even though information may be lost at times after the complete black-hole evaporation
(as, indeed, the original Hawking semiclassical analysis \cite{hawk1} suggests),
the existence of a pure-state description of this information loss avoids 
the use of Hawking formalism \cite{hawk2} in which a state evolves from a pure to a mixed state.
By avoiding this formalism one avoids its pathologies \cite{banks} too.
This makes the information-loss scenario a viable possibility and
there seems to be nothing fundamentally ill about it.

Overall, we see that the pure-state description of black-hole evaporation has a formal meaning,
e.g., by avoiding the formal pathologies discussed in \cite{banks}. At the same time 
its observational meaning is somewhat obscure in the sense that the correlations between
the past inside black hole and future outside of the black hole cannot be measured.
Yet, we believe that even an observational meaning of such correlations is not necessarily
completely hopeless. For instance, the conclusion that such correlations cannot be measured
rests on the assumption that no information can travel faster than light.
Yet, that assumption may not be absolutely true, either because Lorentz symmetry
may be broken at some regime, or particles with a tachyonic dispersion relation may exist.
Indeed, a recent experiment \cite{opera} suggests that muon neutrino travels with a velocity
slightly faster than light. If that will turn out to be true, then one might conceive a method to measure
the above correlations via signals carried by muon neutrinos.

\section*{Acknowledgements}

The author is grateful to an anonymous referee for constructive objections
that stimulated several improvements of the paper.
This work was supported by the Ministry of Science of the
Republic of Croatia under Contract No.~098-0982930-2864.

\appendix

\section{Expansion in terms of inside and outside modes}

Let $x=\{ x^{\mu}\}$ be spacetime coordinates such that $x^0$ is a global timelike coordinate for the whole spacetime. Due to evaporation, we assume that the black hole does not exist for $x^0>t_{\rm evap}$.
Let $\{ \psi_k(x) \}$ be a basis of wave functions for the whole spacetime, which is complete and orthogonal
on any Cauchy hypersurface of constant $x^0$.

Any point $x$ is either inside or outside of the black hole. Therefore $\psi_k(x)$
can be written as a sum of an inside mode and an outside mode
\begin{equation}\label{app1}
 \psi_k(x)=\psi^{( {\rm in})}_k(x) + \psi^{( {\rm out})}_k(x) ,
\end{equation}
where
\begin{equation}\label{app1.1}
\psi^{( {\rm in})}_k(x) \equiv 
\left\{ 
\begin{array}{cl} 
\displaystyle
\psi_k(x)  & {\rm for} \; x \; {\rm inside}, \\
\displaystyle 
0  & {\rm for} \; x \; {\rm outside}, \\ 
\end{array}
\right. 
\end{equation}
\begin{equation}\label{app1.2}
\psi^{( {\rm out})}_k(x) \equiv 
\left\{ 
\begin{array}{cl} 
\displaystyle
0  & {\rm for} \; x \; {\rm inside}, \\
\displaystyle 
\psi_k(x)  & {\rm for} \; x \; {\rm outside}. \\ 
\end{array}
\right. 
\end{equation}
In particular, for $x^0>t_{\rm evap}$, Eqs.~(\ref{app1.1})-(\ref{app1.2}) reduce to a
simpler form
\begin{equation}
 \psi^{( {\rm in})}_k(x) =0, \;\;\;\; \psi^{( {\rm out})}_k(x)=\psi_k(x) ,
\;\;\;\; {\rm for} \; x^0>t_{\rm evap} .
\end{equation}

The most general 1-particle wave function has the form
\begin{equation}\label{app2}
 \psi(x)=\sum_{k}c_k \psi_k(x) = \sum_{k}c_k 
[\psi^{( {\rm in})}_k(x) + \psi^{( {\rm out})}_k(x) ] .
\end{equation}
In particular, if $\psi(x)$ vanishes outside of the black hole, then 
(\ref{app2}) can also be written as
\begin{equation}\label{app3}
 \psi(x)=\sum_{k}c_k \psi^{( {\rm in})}_k(x) .
\end{equation}
Note that, inside the black hole, the modes $\{ \psi^{( {\rm in})}_k(x) \}$ are not necessarily orthogonal
and may be overcomplete. 
Yet, the expansion (\ref{app3}) is well defined
as an expression derived from (\ref{app2}), which, in turn, is an expansion in terms of
modes $\{ \psi_k(x) \}$, which are complete and orthogonal on Cauchy hypersurfaces for 
the whole spacetime.

The most general 2-particle wave function has a form
\begin{equation}\label{app4}
 \psi(x_1, x_2)=\sum_{l}\sum_{k}c_{lk}\psi_l(x_1) \psi_k(x_2) .
\end{equation}
Using (\ref{app1}), this can be written as
\begin{eqnarray}\label{app5}
 \psi(x_1, x_2) & = & \sum_{l}\sum_{k}c_{lk} 
[ \psi^{( {\rm in})}_l(x_1) \psi^{( {\rm in})}_k(x_2) 
+ \psi^{( {\rm out})}_l(x_1) \psi^{( {\rm out})}_k(x_2)  
\nonumber \\
& & \;\;\;\;\;\;\;\;\;\;\;\;
+ \psi^{( {\rm out})}_l(x_1) \psi^{( {\rm in})}_k(x_2)
+ \psi^{( {\rm in})}_l(x_1) \psi^{( {\rm out})}_k(x_2) ] .
\end{eqnarray}
For such a general wave function it may be that both particles are inside, or both
particles outside, or one particle inside and one particle outside.
However, if the particles are created by black-hole evaporation, then
it is always the case that one particle is inside and the other outside. 
For such a more restrictive case
the coefficients $c_{lk}$ in (\ref{app5}) are not completely arbitrary.
For instance, if $c_{lk}\neq 0$ only for those $l,k$ for which
$\psi^{( {\rm out})}_l(x)=0$ and $\psi^{( {\rm in})}_k(x)=0$,
then only the last term in (\ref{app5}) contributes, so 
(\ref{app5}) reduces to
\begin{equation}
 \psi(x_1, x_2)=\sum_{l}\sum_{k}c_{lk} \psi^{( {\rm in})}_l(x_1) \psi^{( {\rm out})}_k(x_2) .
\end{equation}
This is Eq.~({\ref{E6}), here obtained as a special case of  ({\ref{app4}).

\end{document}